# Online Social Networking Has a Greater Effect on Others than on Me: A Third-Person Effect Perspective

**Alireza Heravi**
School of Information Technology and Mathematical Sciences
University of South Australia
Adelaide, Australia
Email: alireza.heravi@mymail.unisa.edu.au

**Sameera Mubarak**
School of Information Technology and Mathematical Sciences
University of South Australia
Adelaide, Australia
Email: sameera.mubarak@unisa.edu.au

**Kim-Kwang Raymond Choo**
School of Information Technology and Mathematical Sciences
University of South Australia
Adelaide, Australia
Email: raymond.choo@unisa.edu.au

## Abstract

To date, much research has been conducted on the positive and negative effects of online social networking (OSN). However, how users perceive others and themselves being subject to these effects and the consequences of users' perceptions are understudied. Drawing from the third-person effect theory, this study examines the self-other perceptual gap for positive and negative effects of OSN and the consequences of perceptions for negative effects. Findings from our online survey (N=187) and interviews (N=8) suggested a significant difference between the perceived positive and negative effects on self and on others. Furthermore, the link between the third-person perception for usage risks of OSN and support for taking privacy protection actions was confirmed. We also found that the self–other discrepant perceptions were not influenced by age, time spent on OSN, number of OSN friends. However, gender emerged as a key difference in the third-person effects gap for privacy risks.

## 1 Introduction

Online social networking (OSN) refers to "web-based services that allow individuals to (1) construct a public or semi-public profile within a bounded system, (2) articulate a list of other users with whom they share a connection, and (3) view and traverse their list of connections and those made by others within the system" Boyd and Ellison (2007, p. 211). With over one billion people worldwide using OSN, the potential benefits and risks of using OSN have been examined in various studies (e.g. Christy et al. 2015; Debatin et al. 2009; Ellison et al. 2007; Krasnova et al. 2010; Reinecke and Trepte 2014). However, it is not clear how OSN users perceive themselves and others to be likely to experience these potential benefits and risks, and what the implications of these perceptions are. In other words, the gap between users' perceptions on the benefits and risks of using OSN on "self" and on "others" is not fully investigated. The current study is designed to examine these self-others discrepancies and its implications.

A common approach to explore the self-others discrepancies with reference to media impact is to use the third-person effect theoretical framework (Paul et al. 2011; Perloff 1993). The theory posits that (1) people perceive others to be more influenced by media than themselves, and (2) as a result of such perceptions their attitudes or behaviours may change (Davison 1983; Gunther and Storey 2003). Although the third-person effect has been used to study various media effect , very few studies have applied the theory in the context of OSN.

Schweisberger et al. (2014) employed the third-person effect approach to investigate the discrepancies concerning the effects of news stories in Facebook on "self" and "others". The results suggested that while participants showed third-person effect for "Low-Relevance" news stories, they did not perceive "High-Relevance" news stories as having a greater effect on others than on themselves. Zhang and Daugherty (2010) studied the third-person effect of American and Chinese OSN users. Their findings indicated that users in both countries believed OSN had more effect on others than on themselves. Furthermore, the OSN users estimated that OSN has a greater influence than TV and print media on other users compared to themselves. They also found evidence for the behavioural component of the third-person effect; those who exhibited a higher degree of third-person effect were less likely to visit an OSN recommended by their friends.

In another study of third-person effect in the OSN context, Paradise (2012) explored the perceived negative effects of Facebook use on "self" and "others" regarding to personal relationships, future employment opportunities, and privacy. Their study showed users believed that using Facebook has a larger negative impact on others than on themselves. In addition, the third-person effect on future employment of younger Facebook users predicted the support for strict regulations protecting privacy in Facebook.

The aforementioned studies demonstrated evidence for the third-person effect in the context of OSN. However, the studies are limited. For example, Zhang and Daugherty (2010) investigated the general effects of OSN while Paradise (2012) studied the negative effects of Facebook, yet neither study examine both the positive and negative effects of OSN. To the best of our knowledge, no research has examined the benefits of using OSN from the third-person effect perspective. Furthermore, the impact of the third-person effect for OSN risks on users' information privacy concerns is not fully investigated. To fill this research void, the current study will examine the gap between "self" and "others" perceptions on the benefits and risks of using OSN. Additionally, the impact of the perceptual gap for the risks of using OSN on the support for using privacy protective actions and on users' information privacy concerns will be investigated.

## 2 Literature Review

### 2.1 Third-Person Effect

The third-person effect hypothesis, proposed by Davison (1983), posits that people perceive mass media as having a greater impact on others than on themselves (the perceptual component). Furthermore, these perceptions may lead people to take some actions (the behavioural component) such as voting, censorship, and purchasing. This notion is named the third-person effect because: "In the view of those trying to evaluate the effects of a communication, its greatest impact will not be on "me" [the first person] or "you," [the second person] but on '"them"-the third persons" (Davison 1983, p. 3). Figure 1 illustrates the notion of the third-person effect.

In the literature, the dichotomy between the perceived effects of mass media on the "self" and "others" - the perceptual component - has received significant attention (e.g. Gunther and Storey 2003; Innes and Zeitz 1988; Paradise 2012; Paul et al. 2000; Perloff 1993; Perloff 1999; Salwen and Dupagne 1999; Zhang 2013). Due to the anticipated effect of the media on others, people may demonstrate attitudinal or behavioural reactions; that is the behavioural component (Davison 1983; Perloff 1999). The literature has focused on support for censorship to examine the behavioural component of the third-person effect (Gunther and Storey 2003). Existing studies suggest that while there is support for censorship of subjects like (child) pornography and violent television programmes, censorship of news or political media content is not fully supported.

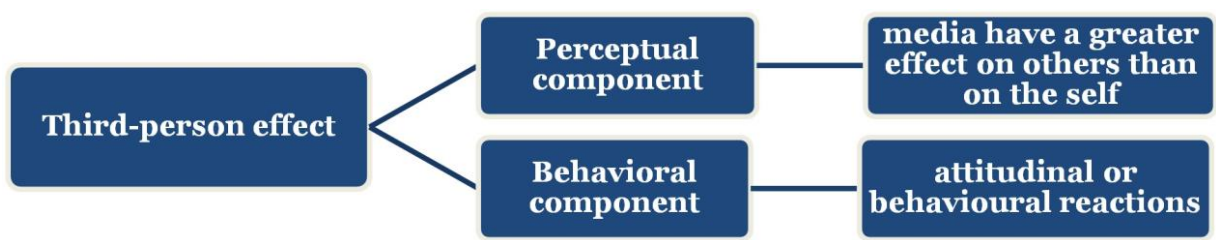

*Figure 1: The third-person effect*

The behavioural component, however, is not limited to censorship. The focal point in the second component of the third-person effect is the changes in attitudes and behaviours that are due to the gap between "self" and "others" perceptions on the effects of the media. Therefore, to investigate the changes in attitudes, this study examines OSN users' concern about information privacy and their support for using privacy protection actions. Compared to the perceptual component, the evidence of behavioural component in the literature is sparser (Gunther and Storey 2003). This partly motivates this research – to examine the behavioural component in the OSN context.

## 2.2  Usage and Privacy Risks in OSN

Information privacy risks exist whenever personal or sensitive information is collected and stored. OSN provides an information-rich environment for cyber criminals (Baltazar et al. 2009; Khan and Mashiane 2014; Krubhala et al. 2015) and third parties (Debatin et al. 2009; Hazari and Brown 2013; Wang et al. 2011). Privacy risks includes personal information being accessed by companies/government agencies (Harold Abelson et al. 2015; Lucas and Borisov 2008; Purtova 2015), or individuals you do not want (Johnson et al. 2012). Added to privacy risks, there are other risks associated with using OSN such as online stalking (Niland et al. 2015) and gossip or spreading rumours (Bumgarner 2007; Debatin et al. 2009). To examine the third-person perception on usage and privacy risks in OSN, we propose the following hypotheses:

*H1a: OSN users will perceive other users being more likely to be exposed to usage risks.*

*H1b: OSN users will perceive other users being more likely to be exposed to privacy risks.*

Furthermore, if OSN users perceive other users are more likely to be exposed to negative outcomes of practicing OSN, this perception may trigger their support for using protective actions, such as providing less personal information and using strict privacy settings. Hence, to explore the behavioural component of the third-person effect in OSN context we propose the following hypotheses:

*H2a: Third person effects for OSN usage risks will increase the support for using protective actions.*

*H2b: Third person effects for OSN privacy risks will increase the support for using protective actions.*

## 2.3  Information Privacy Concern in OSN

Previous research found that users are more likely to disclose personal information on OSN than on non-OSN-mediated communication method (Acquisti and Gross 2006; Fogel and Nehmad 2009). Moreover, OSN users claim to understand information privacy threats but they disclose large amounts of personal information partly due to perceiving that the benefits outweigh the risks (Debatin et al. 2009). However, the OSN users who experienced information privacy breaches were more cautious in sharing personal information (Hazari and Brown 2013). The question that remains to be answered is whether third-person perceptions of OSN usage and privacy risks influence information privacy concerns. Hence, the following hypotheses are proposed:

*H3a: Third person effects for OSN usage risks will influence information privacy concerns.*

*H3b: Third person effects for OSN privacy risks will influence information privacy concerns.*

## 2.4  Benefits of Using OSN

To examine the self-other perception on the benefits of using OSN, we explored peoples' motives for utilising OSN. The connection between motivation and benefits can be explained through the uses and gratifications (Katz et al. 1973) theory, which is a framework for examining the motives for media use. The theory is user-centred and posits users select the media that fulfil their needs (Blumler 1979; McQuail 1994). Therefore, users are motivated to use a medium that gratifies them. In other words, users' motives for using a medium derive from their needs and when their needs are fulfilled they in fact would benefit. For example, consider an individual that is motivated to use OSN to maintain relationships. By using OSN the individual manages his/her current relationships and, therefore, the motive for using OSN would be same as the benefit that he/she would receive. However, it can be argued that motivation may not always lead to benefit. For example, a user who mainly uses OSN for entertainment may face a negative consequence, such as having his/her profile hacked. It has to be noted, however, the current study examines the perceived benefits of OSN rather than the actual positive outcomes. According to Ruggiero (2000, p. 15), benefit is one of the "user-oriented dimensions of interactivity" that is useful to consider in uses and gratifications research.

Based on our review of the literature, we adopted the four most common motives identified in previous studies (e.g. Ancu and Cozma 2009; Joinson 2008; Park et al.'s 2009), namely, 1- entertainment (using OSN for fun), 2- relationship maintenance (managing current relationships), 3- relationship building (establishing new relationships), and 4- information seeking (find out about others).

In the third-person effect literature, it is believed that when the message content is perceived to be negative, people tend to feel that others will be more influenced than themselves and when the message content is perceived to be positive the third-person perception does not occur (Rojas et al. 1996). It is, therefore, interesting to examine whether OSN users exhibit third-person perception about the benefits\motives for using OSN. Based on the above argument, we propose the following hypotheses:

*H4(abcd): OSN users will show third-person perception for (a) entertainment, (b) relationship maintenance, (c) relationship building, and (d) information seeking.*

## 3  Research Methodology

In this study, both quantitative and qualitative research methods were employed. We used purposive sampling method in recruiting research participants. It was not possible to apply random sampling in quantitative research as the complete list of the studied population (students enrolled in our case study university) was not available to the authors. In addition, in qualitative research, random sampling is not generally required (Marshall 1996; Ritchie et al. 2013). This is due to the fact that the underlying principle for selecting interviewees is often derived from the research questions and it is acknowledged that some participants are more likely to provide more insights than others.

Participants for both the survey and the interview received an information sheet explaining the aim and scope of the research. In addition, interview participants (interviewees) signed a consent form before being interviewed. No personal information was asked in the survey or the interview. Ethics approval was obtained from the Human Research Ethics Committee of the University of South Australia prior to the start of this research.

The online survey was advertised through an announcement at the university's student portal for two weeks in July 2015. We received 188 valid responses. From the online survey respondent pool, eight participants (four female, four male) were selected for face-to-face interviews. It is expected that experienced OSN users are more likely to provide insights than less experienced ones. In other words, the selection criteria for interviewees were 1-being a member of an OSN for at least two years and 2-using OSN on a daily basis for at least 30 minutes. The semi-structured interviews lasted on average 30 minutes and they were recorded, transcribed, and then analysed. Using a deductive approach, the transcripts were read line-by-line and prominent themes were identified. Then, the catalogue of prominent themes were analysed to find the similarities and differences.

### 3.1  Survey Participants

The majority of the survey respondents (n=188) were male (52.1%, n=98), Australian (57.9%, n=109) and close to half of the respondents (49.4%, n=93) were aged 19-25. On average, participants spent 91.06 minutes per day (Max=300, Min=5) on OSN and had 104.53 OSN friends (Max=2554, Min=4). More than half of the survey respondents (52.65%, n=99) admitted that they do not know at least half of their OSN friends. The highest education level of most of the respondents (73.4%, n=138) was either a high school certificate (n=73) or a Bachelor's degree (n=65).

### 3.2  Survey Measures

Previous research has indicated that the order of the questions (self-then-others versus others-then-self) does not affect the third-person effect (Dupagne et al. 1999; Price and Tewksbury 1996). Moreover, the back-to-back questions that immediately contrast the effects on "self" and "others" also do not have a significant effect on the third-person effect (Price and Tewksbury 1996) and many of the studies in this field have employed this questioning format (e.g. Banning and Sweetser 2007; Innes and Zeitz 1988; Zhang and Daugherty 2010). Therefore, in the online survey, we employed the back-to-back self-then-others approach.

To examine the difference between perceived first (self) and third (others) person effects, participants were asked to indicate separately how they perceived the motivations for using OSN and the risks of using OSN had an effect on themselves and other users. As listed in Table 1, the scales used for this research consist of three items and the wording of the items for "self" and "others" was identical except for the first or third person pronouns. Each scale was presented as a multiple choice question with items having the same semantic value. The participants could select no item, only one or more than one items. The selection of one item did not affect the selection of other items. However, to avoid bias, the order of the scale items was randomised for each participant. For a selected item, a score of 1 was assigned, and for a non-selected item, a score of 0 was given. Therefore, the scores of each scale range from 0 (no selection) to 3 (all three selected). Subsequently, to measure the third-person perception on each topic, "self" scores were subtracted from the "others" scores. It has to be noted that in the third person effect studies, the interest is on the score of the scales rather than on an individual item within the scales. Therefore, all items had the same semantic value.

| **Entertainment** (developed by authors) |
| --- |
| 1-I(people) have fun interacting with others through OSN.<br>2-I(people) enjoy using OSN.<br>3-Using OSN provides me(people) with a lot of enjoyment. |
| **Relationship maintenance** (Joinson 2008; Papacharissi and Mendelson 2011) |
| 1-I(people) use OSN to keep in touch with my(their) friends/family.<br>2-I(people) use OSN to maintain relationships with people I may not get to see very often<br>3-I(people) use OSN to find people I(they) haven't seen for a while. |
| **Relationship building** (Foregger 2008; Park et al. 2009) |
| 1-I(people) use OSN to make new friends.<br>2-I(people) use OSN to network with new people.<br>3-I(people) use OSN to meet interesting people. |
| **Information seeking** (developed by authors) |
| 1-By using the OSN I(people) find out what's going on.<br>2-By using the OSN I(people) become aware of what my(their) friends are up to.<br>3-By using the OSN I(people) learn more about others. |
| **Perceived privacy risks** (Dinev and Hart 2004; Krasnova and Veltri 2010) |
| 1-Personal information be misused.<br>2-Personal information be made available to unknown individuals, companies or government agencies without your knowledge.<br>3-Personal information be accessed by someone you don't want (e.g "ex", parents, teacher, employer, etc.) |
| **Perceived usage risks** (Debatin et al. 2009) |
| 1-Unwanted advances, stalking, or harassment.<br>2-Damaging gossip or rumours<br>3-Personal data stolen/abused by others. |
| **Protective actions** (developed by authors) |
| 1-OSN users should provide very limited personal information in their profile.<br>2-OSN users should provide less information about themselves in their wall, comments, etc.<br>3-OSN users should use strict privacy settings even if this results in only a few people having access to the user's page. |
| **Information privacy concern** (Malhotra et al. 2004) |
| 1-By using OSN, I am concerned about threats to my personal privacy.<br>2-Compared with other subjects on my mind, personal privacy in OSN settings is very important.<br>3-Compared to others, I am more sensitive about the way OSN providers handle my personal information. |

*Table 1. Survey measures*

We used existing measures, wherever possible, although we changed the wordings of some of the adapted measures for this study. The final survey consists of nine sections. In the first section, demographic questions were asked. In addition, participants indicated their daily average time spent on OSN and their number of OSN "friends". As listed in Table 1, the remaining sections of the survey were designed to collect data regarding perceived benefits\motives for using OSN, perceived privacy and usage risks, protective actions and information privacy concern.

# 4 Results
## 4.1 Quantitative Findings

Hypotheses 1a and 1b predicted that people would perceive (a) usage risks and (b) privacy risks of OSN will more likely be experienced by other users than themselves. Data analysis supported these hypotheses (see Table 2). Two paired-samples t-test revealed that mean perceived effect for usage risks and privacy risks on "others" was higher than the mean perceived effect of the two topics on self. Therefore, respondents indeed believed other OSN users are more at risks than themselves. To further examine the third-person perception, frequency analysis was conducted. The findings revealed that 61.4% (n=115) of the participants perceived the usage risks of OSN is greater for others than themselves, while only 1.1% (n=2) considered themselves to be more at risk than other OSN users. Similarly, 48.6% (n=91) believed the privacy risks of OSN are more for others and only 3.2% (n=6) perceived the opposite to be the case.

|  | Mean "others" | Mean "self" | Test statistics* |
|---|---|---|---|
| Usage risks | 2.556 | 1.289 | ($t$=15.112, $df$=186, $P$<.001) $md$=1.26 |
| Privacy risks | 2.535 | 1.861 | ($t$=9.043, $df$=186, $P$<.001) $md$= 0.67 |
| Entertainment | 1.583 | 1.150 | ($t$=6.642, $df$=186, $P$<.001) $md$= 0.43 |
| Relationship maintenance | 2.26 | 1.749 | ($t$=8.188, $df$=186, $P$<.001) $md$= 0.51 |
| Relationship building | 1.877 | .647 | ($t$=13.444, $df$=186, $P$<.001) $md$= 1.22 |
| Information seeking | 2.358 | 1.807 | ($t$= 7.408, $df$=186, $P$<.001) $md$= 0.55 |

Note:* Two-tailed t-test, degree of freedom, $p$ value and mean differences. Means reflect values on a scale from 0= no effect at all to 3= a great deal of effect.

*Table 2. Mean differences for "self" and "others"*

|  | 1 | 2 | 3 | 4 | 5 |
|---|---|---|---|---|---|
| 1-Information Privacy concern | 1 |  |  |  |  |
| 2-Usage risks "self" | .220** | 1 |  |  |  |
| 3-Usage risks "others" | .171* | .280** | 1 |  |  |
| 4-Privacy risks "self" | .246** | .399** | .316** | 1 |  |
| 5-Privacy risks "others" | .226** | .175* | .679** | .431** | 1 |

Note: ** $p$<.01, * $p$<.05.*

*Table 3. Correlation between OSN risks and privacy concerns*

To address H2a, which predicted the third-person effect for usage risks would increase the support for using protective actions, a linear regression was conducted. The result (F(1,185)=4.852, p<.05) indicated the third-person effect for usage risks of OSN could statistically predict protective actions. Further, the correlation analysis demonstrated that the third-person effect for usage risks of OSN was positively correlated with the protective action (r(187)=.160, p<.05).

H2b, that predicted third-person effect for OSN privacy risks would increase the support for using protective actions, was not supported. The two variables were not correlated.

Hypotheses 3a, and 3b, which predicted third-person effect for (a) usage risks and (b) privacy risks would influence information privacy concerns, were not supported. Correlation analyses showed that self-other perception gap for OSN usage risks and privacy risks were not correlated with information privacy concerns. However, further analyses (see Table 3) indicated that usage risks for "self" and "others" and privacy risks for "self" and "others" were positively correlated with information privacy concern. Therefore, perceived higher usage and privacy risks for "self" and "others" is associated with being more concerned about information privacy.

Hypotheses 4a, 4b, 4c and 4d, which predicted people would display self-others perception for (a) entertainment, (b) relationship maintenance, (c) relationship building and (d) information seeking, were supported. Four paired-samples t-test showed (see Table 2) that the mean perceived effect for the benefits\motives for "others" was higher than for "self". To further examine the degree to which the third-person perception occurred, frequency analyses were conducted. As shown in Table 4, relationship building had the highest third-person perceptual gap followed by entertainment. Both of these benefits\motives had the same value for first-person effect.

|  | Greater effect on "others" | Greater effect on "self" | Same effect |
| --- | --- | --- | --- |
| Entertainment | 49.1% (n=92) | 4.8% (n=9) | 46% (n=86) |
| Relationship maintenance | 47.6% (n=89) | 3.8% (n=7) | 48.7% (n=91) |
| Relationship building | 62.5% (n=117) | 4.8% (n=9) | 32.6% (n=61) |
| Information seeking | 47.6% (n=89) | 4.3% (n=8) | 48.1% (n=90) |

*Table 4. Frequency analyses*

To examine possible conditions that facilitate or hinder the third-person effect, correlations between age, time spent on OSN, number of OSN friends, and third-person perceptions were tested. No significant correlations were observed.

To explore if differences in the third-person effect gap was due to gender, independent samples t-test was conducted. The mean difference between women and men was only significant for the privacy risks gap ($t(185)=1.394$, $p=003$). Further analysis revealed that men perceived using OSN is associated with higher privacy risks than women. About twice the number of women (12.3%, n=11), men (26.5%, n=26) scored 2 or higher (on a scale of 0=no privacy risks to 3=high privacy risks) in privacy risk scale.

A series of one-way ANOVA were conducted to determine if the third-person effects were different for participants according to education level. The self-other perception gaps for relationship maintenance ($F(4, 182)=4.269$, $p<.005$) and relationship building ($F(4, 182)=2.959$, $p<.05$) were statistically different for different levels of education. There was an increase in the self-other perception gap for relationship maintenance from 0.61 for high school graduate to 1.57 for PhD students, an increase of 0.955, which was statistically significant ($p=.011$). Similarly, the gap for relationship building increased from 1.31 for high school graduate to 1.71 for PhD students, an increase of 0.399, which was not statistically significant ($p=.387$).

## 4.2 Qualitative Findings

From the qualitative analysis, three main topic areas emerged: 1- third-person perception about the risks of using OSN- the discrepancy between the perceived risk of using OSN on "self" and "others", 2- third-person perception about the benefits of using OSN- the discrepancy between the perceived benefits of using OSN on "self" and "others", and 3- privacy protection behaviour- support for taking protective actions based on third-person perception.

### 4.2.1 Third-Person Perception: The Risks of Using OSN

All interview participants perceived other OSN users are more at risk than themselves. They were also concerned about their information privacy and believed using OSN involves potential risks such as, personal information (including location) misused or accessed by someone they did not want. One interview participant who was an experienced OSN user was concerned about the collective users' information available on OSNs. He believed other OSN users are more at risk because "they are not aware how much the risks can cost them. It might cost them their marriage, job and other unnecessary scams and disturbance".

The reasoning of three interviewees for exhibiting third-person perception towards the risks of using OSN were: 1- understanding the consequences of the risks of using OSN, 2- being familiar with the privacy setting of their OSN, and 3- properly utilising the privacy setting. In other words they claimed to be thoughtful and knowledgeable in managing their information privacy while generally other users do not have these qualities. One of these interviewees said:

> privacy protection is all about awareness. I think all OSN users face the same risks.
> But I'm sure that I'm more aware of privacy issues and I'm more careful than others.
> This is the only difference between me and others.

These three participants were all active Facebook users and have used the website for more than seven years. This is in line with other studies (Gunther 1995; Rucinski and Salmon 1990; Tiedge and Silverblatt 1991), whose findings indicate people with education qualifications are more prone to the third-person effect.

Five other interviewees justified their perception about the risks of OSN by stating: 1-other users share astonishing amount of personal information, 2- other users are careless about securing their information, and 3- other users spend more time on OSN.

Overall, all interview participants displayed high third-person perception and perceived "others" are more likely to face negative consequences of using OSN than themselves. This is in line with the findings of Kim and Hancock (2015) who confirmed that Facebook users perceived others are more likely to experience negative outcomes of using it than themselves.

### 4.2.2 Third-Person Perception: The Benefits of Using OSN

Most of the interviewees (n=7) perceived the benefits of using OSN are more for other users than themselves. This echoes the findings of Kim and Hancock (2015). The interviewees presented a range of reasoning for their perceptions. Two interviewees believed people who use OSN for business benefit more. One of these interviewees said: "the benefit depends on how you use social media. For instance, I know a friend who has a page on Facebook and sells some products. Compare to him I benefit less. I just kill time." Age was another factor that was considered by an interviewee. She believed younger users benefit more than older users. It was interesting that this interviewee who is 32 years old considered herself to be 'old' for a Facebook user.

Two interviewees asserted that the range of services used is related to the amount of benefit the OSN users receive. One of them stated, "definitely the benefits of using Facebook for me and others are not the same because I don't use many of the services that Facebook provides. I consider myself as a minimalist regarding using Facebook. I mainly look at the news feeds". Other participants believed that other users use OSN as a medium to communicate with others and, therefore, benefited more. However, all these interviewees reported the main benefit of practicing OSN for them is to stay in contact with friends/families.

All the interview participants believed OSN has several benefits. When they were asked to name the most beneficial feature of OSN, four stated staying in contact with friends/families, while another two said seeking information and one specified passing time. It has been discussed in the literature that users' desire to keep in touch with friends is the most common motive for using OSN (Wilson et al. 2012).

### 4.2.3 Third-Person Behaviour: Privacy Protection

There was a clear connection between interview participants' third-person perception about the risks of using OSN and their support for taking privacy protection actions. All interview participants believed that OSN users should provide very limited personal information and use strict privacy settings. Interestingly, interview participants' attitudes toward privacy protection actions and their privacy behaviour were the same. This contradicts findings of Debatin et al. (2009) who suggested that peoples' concerns about privacy are not necessarily linked to their privacy behaviour.

## 5 Discussion

In the current study, we examined several constructs to examine the third-person effect in the OSN context. H1a and H1b predicted that people would perceive (a) usage risks and (b) privacy risks of OSN are more likely to be experienced by other users than themselves. Both hypotheses were supported. This is in line with the findings of Debatin et al. (2009) who concluded Facebook users ascribed the risks to privacy more to others than to themselves. Similarly, Paradise's (2012) results suggested that Facebook users displayed third-person perception about the negative impact of Facebook use regarding personal relationships, future employment opportunities, and privacy.

Regarding hypotheses H2a and H2b, which predicted the third-person effect for (a) usage risks of OSN and (b) OSN privacy risks would increase the support for using protective actions, only H2a was supported. The support for H2a echoes the findings of Kim and Hancock (2015) who showed Facebook users who believed other users are more likely to be exposed to negative social outcomes (e.g. victim of cyberbullies, target of scams) supported the protective action regulations. The result for H2b may be explained by the fact that the self-others perception gap for privacy risks (0.67) was about half the gap for usage risks (1.26), indicating fewer participants perceived other users are more likely to be exposed to privacy risks (relative to usage risks) than themselves. Another possible reason for this outcome can be that privacy risks, compared to usage risks, were perceived as less amenable to enacting protection behaviours. Perhaps participants believed that usage risks are more serious than privacy risks. The mean difference between women and men in privacy risks gap was significant. More men exhibited third-person effect about privacy risks than women. Gender's impact on the self–other perceptual gap has been reported in other studies. For example, Zhang and Daugherty (2010) demonstrated American OSN users displayed a greater third-person effect on the impact of OSN for "other" females than "other" males.

Hypotheses 3a, and 3b, which predicted the third-person effect for (a) usage risks and (b) privacy risks will impact information privacy concerns, were not supported. However, perceived usage risks and perceived privacy risks for both "self" and "others" were positively correlated with information privacy concern. This indicates that concern about information privacy is not necessarily related to individuals' third-person perception.

Hypotheses 4a, 4b, 4c and 4d, which predicted users would exhibit third-person perception for (a) entertainment, (b) relationship maintenance, (c) relationship building and (d) information seeking, were supported. This indicates the self–other discrepant perceptions toward the benefits of using OSN. This is consistent with Kim and Hancock's (2015) findings where it was suggested that Facebook users assessed other users as being more likely to encounter positive Facebook outcomes (e.g. more social support and a better sense of connectedness to the community) than themselves. Interestingly, the participants displayed third-person effect about the possible negative and positive effects of OSN. This finding is in contrast with previous studies that suggested when the impact of the media is perceived to be positive people will show first-person effect (Rojas et al. 1996). It appears that OSN users tend to believe OSN has much to offer but since they only use a small fraction of what is offered they are not benefiting as much as other users are. Perhaps OSN users perceive other users access more features of OSN and therefore reap the benefits.

The key limitation of this study is the relatively small sample size and the use of university students. This limits the generalizability of the findings to the general public, although we observe that the use of university student sample appears to be the norm in a number of studies, including those involving OSN. Therefore, future research would benefit by expanding the scope of the sample population to obtain a more representative sample. Despite the limitation, the study has provided insights into OSN in the context of third–person effect, which contributes to the media effects literature, as well as improving the user's understanding of the self–other discrepant perceptions in using OSN. It is hoped, for example, participants of this research will consider about the potential security and privacy concerns and undertake protection measures in their future OSN activities.

## 6  Conclusion

The current study utilised both quantitative and qualitative methods to examine the third-person effect in OSN. The perceptual component of the theory was explored by examining the self-other perceptual gap for both benefits and risks of using OSN. The findings from the qualitative interviews confirmed the survey findings that OSN users perceived other users are more likely to experience the positive and negative outcomes of OSN use than themselves. These self–other discrepant perceptions were not hindered by age, time spent on OSN, or number of OSN friends. The descending order of the third-person perceptual gap for the benefits and risks of OSN use was found to be: 1- usage risks, 2- relationship building, 3- privacy risks, 4- information seeking, 5- relationship maintenance, and 6- entertainment. The behavioural component of the theory was investigated by examining the impact of the third-person perception gap for the usage and privacy risks of OSN on the support for taking protective action(s) and on user concern about their information privacy. Only the perception gap for usage risks predicted the support for taking privacy protection actions. Users' information privacy concerns were not determined to be influenced by the perception gap for privacy and usage risks of OSN use.

Future research could include an in-depth study of specific OSN features usage, which would have been influenced by the self-others perceptual gap. Examples of such features are sharing contents (i.e. status, stories, articles, videos, photos, and whereabouts), posting comments, liking or tagging, and joining groups/fan pages.

## Copyright